\documentclass[%
preprint, showpacs,
 amsmath,amssymb,
 aps,
prb,
longbibliography,
 lengthcheck,%
]{revtex4-1}

\usepackage{graphicx}
\usepackage{dcolumn}
\usepackage{bm}
\usepackage{hyperref}

\newcommand{\chiach}{$\chi_{ac}(H)$~}
\newcommand{\chiact}{$\chi_{ac}(T)$~}

\newcommand{\tc}{$T_C$~}
\newcommand{\hcone}{$H_{c1}$~}

\begin{document}

\title{Precursor phenomena at the magnetic ordering of the cubic helimagnet FeGe}



\author{H. Wilhelm}
\affiliation{Diamond Light Source Ltd., Chilton, Didcot, Oxfordshire, OX11 0DE,
United Kingdom}

\author{M. Baenitz, M. Schmidt}
\affiliation{Max Planck Institute for Chemical Physics of Solids,
N\"{o}thnitzer-Str. 40, 01187 Dresden, Germany}

\author{U. K. R\"{o}{\ss}ler, A. A. Leonov, A. N. Bogdanov}
\affiliation{IFW Dresden, Postfach 270116, 01171 Dresden, Germany}



\begin{abstract}
We report on detailed magnetic measurements on the cubic helimagnet FeGe in
external magnetic fields and temperatures
near the onset of long-range magnetic order at $T_C= 278.2(3)$~K.
Precursor phenomena display
a complex succession of temperature-driven crossovers and phase
transitions in the vicinity of $T_C$.
The A-phase region, present below $T_C$ and fields $H<0.5$~kOe, is
split in several pockets. Relying on a modified phenomenological theory for
chiral magnets, the main part of the A-phase could indicate the existence of a
$+\pi$ Skyrmion lattice, the adjacent A$_2$ pocket, however,
appears to be related to helicoids propagating
in directions perpendicular to the applied field.
\end{abstract}


\maketitle

Since the discovery of the mono-silicides of transition metals
\cite{Wever30,Boren33} crystallizing in the $B20$ structure (space group
$P2_13$) the peculiarities of their electronic \cite{Manyala08} and magnetic
\cite{Yu10} properties attract an unabated attention.
The magnetically ordered $B20$-compounds are chiral cubic helimagnets as a
consequence of the broken inversion symmetry and the Dzyaloshinskii-Moriya (DM)
interaction \cite{Dzyaloshinskii64,Bak80}.
The paramagnetic to  helimagnetic transition in these compounds is a
long-standing and unsolved problem that seems to be
related to the specific frustration introduced
by the chiral DM interactions.
Intensive experimental investigations of the archetypical chiral helimagnet
MnSi report numerous physical anomalies along the magnetic ordering transition,
and particularly, indicate the existence of a small closed area in the magnetic
field$-$temperature, ($H,T$), phase diagram, the so-called ``A-phase''
\cite{Kusaka76,Komatsubara77,Kadowaki82,Gregory92,Thessieu97,Lamago06,
Neubauer09,Ishikawa84,Lebech95,Grigoriev06,Grigoriev06a,Muehlbauer09}.
Magnetic neutron diffraction data first suggested a paramagnetic nature of the
magnetic state in the A-phase \cite{Ishikawa84}, but later static magnetic
modulations transverse to the field direction were found \cite{Lebech95}.
The attempts to explain the A-phase by the formation of a specific modulated
phase either with a one-dimensional (1D) modulation ("single"-$q$ helicoids)
\cite{Maleyev10,Grigoriev06,Grigoriev06a} or as "triple-$q$" modulated textures
\cite{Muehlbauer09} are in contradiction to earlier experimental data
suggesting a subdivided A-phase in MnSi \cite{Komatsubara77,Kadowaki82} as well
as a rigorous theoretical analysis \cite{Leonov10,Roessler10}.

Theory has predicted the existence of chiral
precursor states near magnetic ordering \cite{Roessler06,Leonov10}
and specific Skyrmionic textures composed
of string-like localized states \cite{Bogdanov89,Bogdanov94}.
Recent direct observations of specific two-dimensional (2D) modulations in
Fe$_{0.5}$Co$_{0.5}$Si \cite{Uchida06} and FeGe \cite{Uchida08} as well as
isolated and bound chiral Skyrmions in Fe$_{0.5}$Co$_{0.5}$Si \cite{Yu10} and
FeGe \cite{Yu10a} at low temperatures conclusively prove the existence of
Skyrmions in \emph{nanolayers} of cubic helimagnets.
However, the relation between the precursor phenomena
in \textit{bulk} material very close to $T_C$
with the direct microscopic observations at much lower
temperature in \textit{nanolayers} \cite{Yu10a,Yu10}
and the theoretical predictions of stable
$-\pi$-Skyrmions \cite{Bogdanov89,Bogdanov94} is still an
open question \cite{Roessler10}.

In this Letter we present detailed magnetic
ac-susceptibility measurements on
FeGe close to the onset of magnetic order.
FeGe is an important counterpart to MnSi as a cubic helimagnet, because it
behaves magnetically differently \cite{Lundgren70,Lebech89,Pedrazzini07}.
Spin fluctuations are less relevant than in MnSi
as the Fe ion bears a stable
magnetic moment characteristic
for strong band-ferromagnets \cite{Lundgren68}.
Our results conclusively show
that the A-phase region is not a distinct simple phase
with a unique single-$q$ multidomain
or triple-$q$ monodomain state.
Furthermore, the detailed ($H,T$) phase diagram displays
a complex succession
of temperature driven crossovers and phase transitions.
The phase diagram can be interpreted as
a generic example of
solitonic mesophase formation.
Theoretically, these precursor phenomena rely
on chiral localized modulations
which behave as solitonic units
with \textit{attractive} interactions close to
the magnetic ordering transitions \cite{Leonov10,Roessler10}.
Our findings establish the chiral magnetic ordering in non-centrosymmetric
magnets as a new paradigm for unconventional phase transitions,
which are expected to occur similarly in various
systems described by Dzyaloshinskii's theory of
incommensurate phases \cite{Dzyaloshinskii64}.

The ac-susceptibility, $\chi_{ac}$, of single crystalline FeGe \cite{Wilhelm07}
was measured in a Quantum Design PPMS device using a drive coil frequency of
1~kHz and an excitation field of 10~Oe. The single crystal ($m=0.995$~mg) with
almost spherical shape was aligned with the $[100]$ axis parallel to the
external magnetic field.
Field sweeps were recorded such that the measuring temperature was always
approached from 300~K in zero field. For temperature runs the sample was
zero-field cooled.

\begin{figure}
\hspace*{-8mm}
\includegraphics[width=0.625\textwidth]{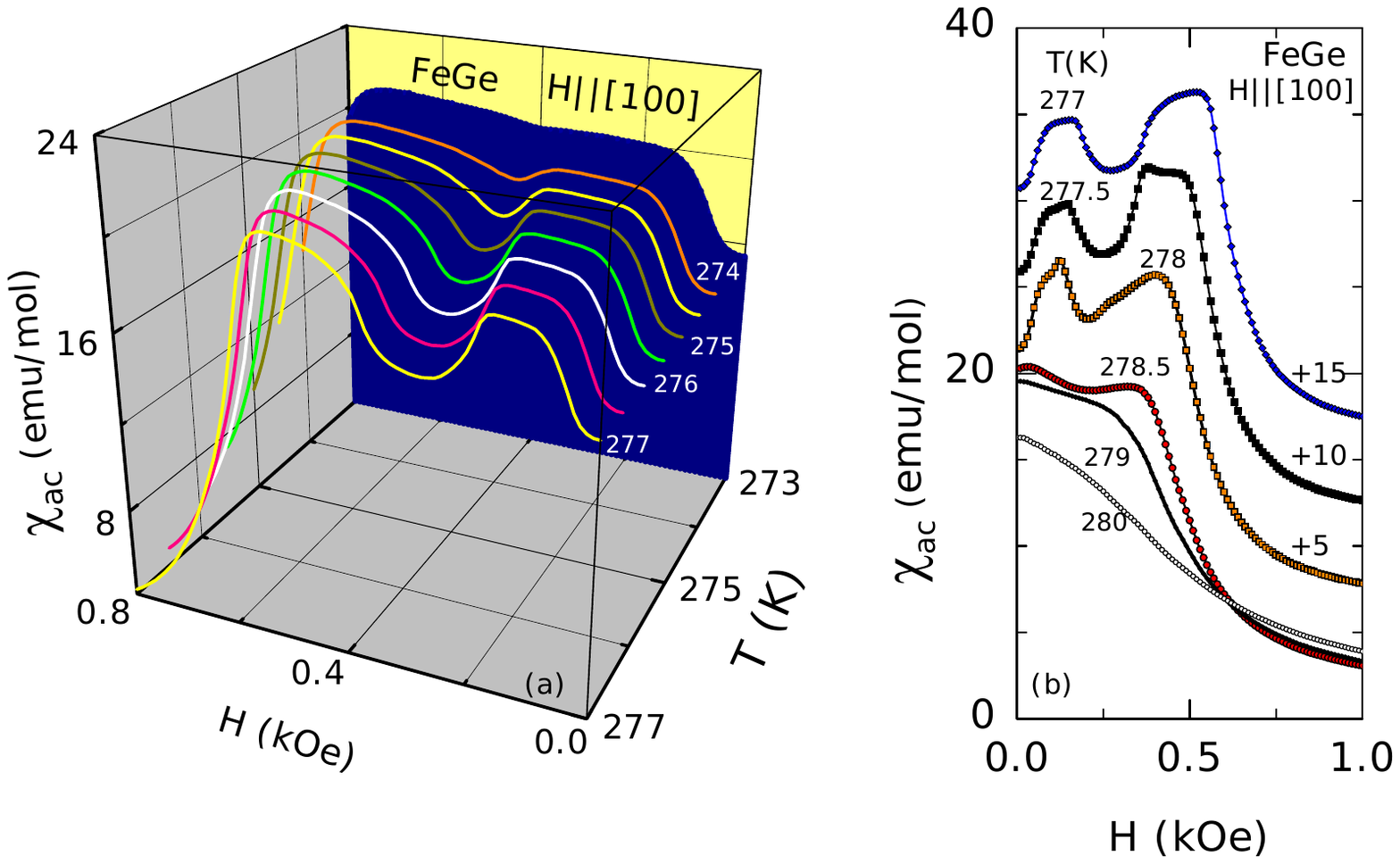}
\vspace*{-100mm} \caption{Isothermal $\chi_{ac}(H)$ data of FeGe for $H
\parallel [100]$. (a) The evolution of a well defined minimum in \chiach upon
approaching \tc from low temperature is the fingerprint of the A-region. (b)
Very close to \tc subtle changes in \chiach below about 0.6~kOe indicate a
complex sequence of magnetic phase transitions and
crossovers.}\label{fig:chilowtzoom}
\end{figure}

Figure~\ref{fig:chilowtzoom} shows the field dependence $\chi_{ac}(H)$ for
temperatures in the vicinity of $T_C=278.2$~K. After an initial increase of
\chiach a minimum starts to evolve at low fields and for temperatures in the
range 272.5~K~$<T<279$~K. This dip in \chiach is well pronounced at 277~K and
ceases as the temperature approaches 273~K (c.f.
Fig.~\ref{fig:chilowtzoom}(a)).
Lower and upper fields $H_{A}^1$,  $H_{A}^2$, delimiting the A-region were
extracted from the inflection points on either side of the minimum. Close to
these fields maxima in the imaginary part, $\chi''_{ac}(H)$, were observed.
Although a shallow minimum is found in \chiach at 278.5~K (see
Fig.~\ref{fig:chilowtzoom}(b)) no maximum in $\chi''_{ac}(H)$ was observed.
Further inflection points below and above at A-region
were used to define the critical fields $H_{c1}$ and $H_{c2}$.
The former corresponds to the entrance
into the cone structure, and the latter is the spin-flip into the field
polarized state.
Near $H_{c1}$ also a peak in $\chi''_{ac}(H)$ was observed.
Cycling the field at 276~K reveals a hysteresis of the order of 20~Oe near all
transition fields $H_{c1} < H_{A}^1 <  H_{A}^2 < H_{c2}$, whereas only a small
hysteresis with width below 10~Oe was observed near $H_{c1}$ and $H_{c2}$ at
275~K.
%
%

\begin{figure}
\hspace*{-3mm}\includegraphics[width=0.575\textwidth]{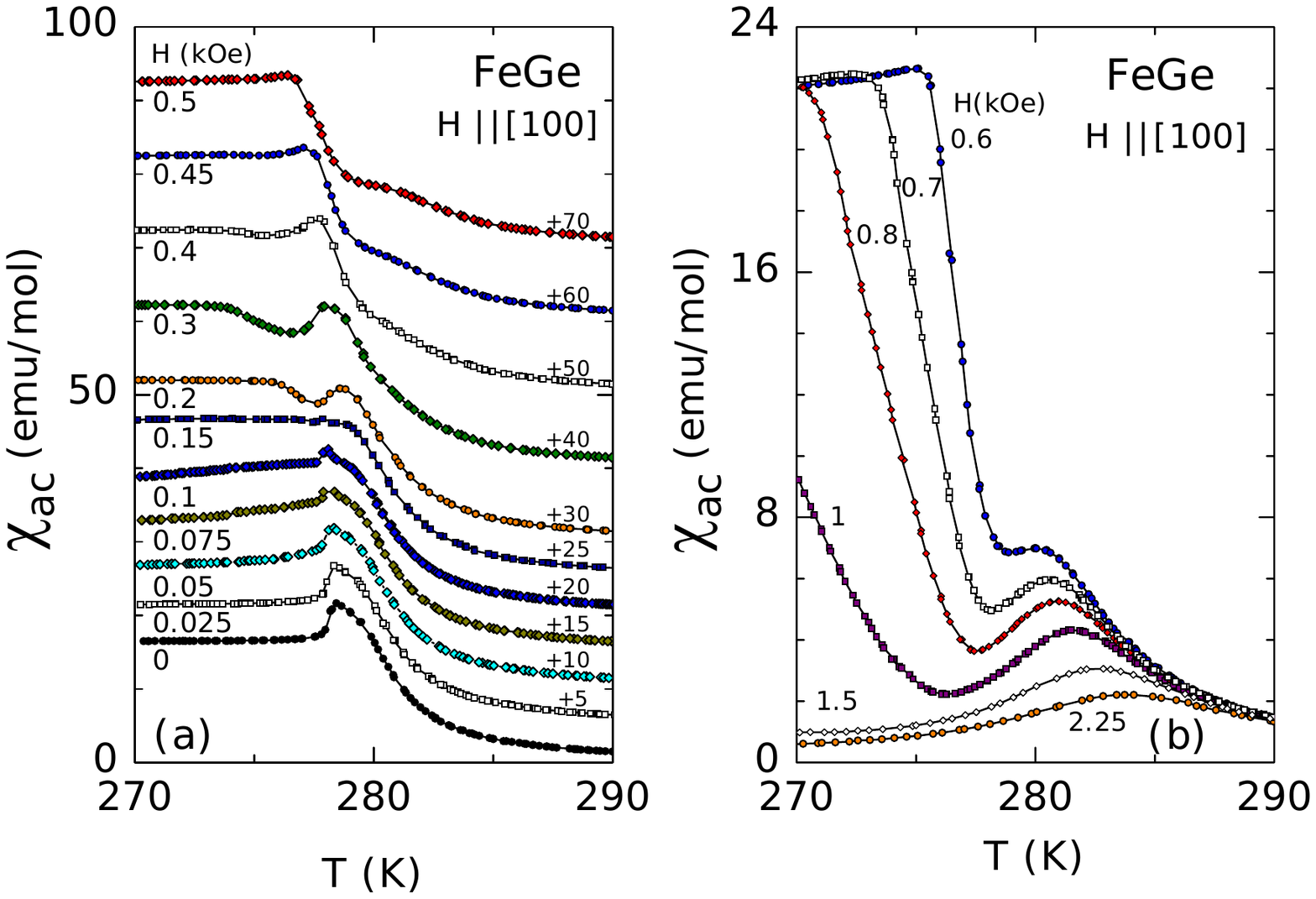}
\vspace*{-87.5mm}\caption{$\chi_{ac}(T)$ of FeGe for constant fields
$H\parallel [100]$. (a) The position of the maximum observed at low fields
coincides with $T_C$. The shallow minimum above $H=0.2$~kOe is due to the
A-region. Curves are vertically shifted. (b) A broad peak in \chiach is clearly
visible for $H>0.5$~kOe. It is related to the crossover from the paramagnetic
to the field-polarized state.}\label{fig:hfa}
\end{figure}

The A-region also appears as a minimum in $\chi_{ac}(T)$ as depicted in
Fig.~\ref{fig:hfa}(a).
Again defining the limits of this region by inflection points, two transition
lines $H_A^L < H_A^H$ were deduced.
Closely above $H_A^H$ we find a strong peak in $\chi_{ac}''(T)$ marking this
anomaly as a possibly discontinuous phase transition.
Importantly, these two transition lines extend
towards higher fields
than the transition line $H_A^2$ found from field runs.

The field region $H \gtrsim H_A^H\approx 0.45$~kOe
displays a characteristic
feature in $\chi_{ac}(T)$ with a shoulder at about 280~K (c.f.
Fig.~\ref{fig:hfa}(a)) which evolves into a clear maximum for $H> 0.6$~kOe as
can be seen in Fig.~\ref{fig:hfa}(b).
This peak broadens and
shifts to higher temperatures with increasing field.
Inflection points below and above this maximum, $H_{\times}^{<}$,
$H_{\times}^{>}$, signal a smeared crossover from the paramagnetic into the
field-polarized state.
Its position reaches 284~K at $H=2.25$~kOe, the highest field measured.
Related to this features is a very shallow minimum in $\chi''_{ac}(T)$ between
$H_{\times}^{<}$ and $H_{\times}^{>}$.

The field values $H_{\times}^{>}(T)$ extrapolate to zero field at $T_0=280$~K.
This implies that an intermediate state exists between $T_0$ and $T_C$.
In an applied magnetic field, this intermediate state transforms
into different magnetically ordered states at lower temperatures.
These transitions are visible in the $\chi_{ac}(T)$ data
(Fig.~\ref{fig:hfa}(a)).
The transition into the helical state is seen as cusps in $\chi_{ac}(T)$ close
to $T_C$ for fields $0 < H < 50$~Oe, while the transition into the conical
helix is seen as a tiny peak (data for $H=100$~Oe).
For fields $H > 100$~Oe, the entrance into the A-region is observed by a
markedly changed behavior of $\chi_{ac}(T)$.
%
%

More details of the magnetic properties in this temperature range are provided
by \chiach depicted in Fig.~\ref{fig:chilowtzoom}(b). Clearly, the overall
field dependence of \chiach changes strongly in the narrow temperature range
277.5~K$\leq T\leq 279$~K and for $H\leq 0.5$~kOe.
The characteristic anomalies present up to 278~K appear completely smeared at
278.5~K where only a shallow minimum is visible between 0.1~kOe and 0.28~kOe.
Only a smooth decay of \chiach remains for $T\geq 279$~K which indicates a
crossover towards the field-aligned state by an inflection point around
0.4~kOe.
Correspondingly, in \chiact data we observe a similar crossover for
0.2~kOe~$<H<0.35$~kOe for temperatures in the same range 279 to 280~K
(Fig.~\ref{fig:hfa}(a)).
%

\begin{figure}
\hspace*{-17.5mm}
\includegraphics[width=0.675\textwidth]{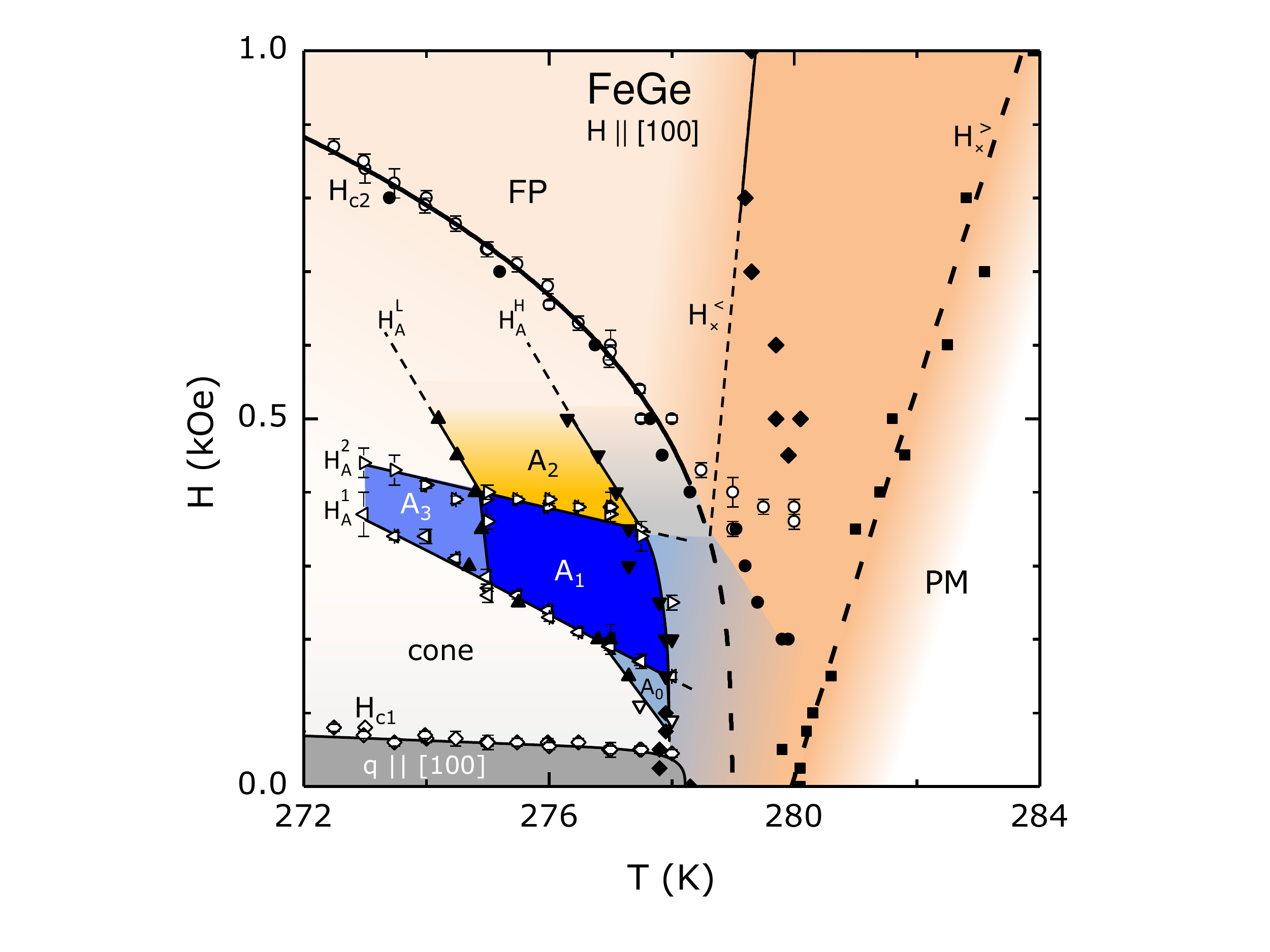}
\vspace*{-5mm}\caption{$(H,T)$-phase diagram of FeGe for magnetic fields
$H\parallel [100]$. Open and bold symbols represent data obtained from \chiach
and $\chi_{ac}(T)$, respectively. Below $T_C=278.2$~K various phases are
observed: a helical state with $q\parallel [100]$ ($H<H_{c1}$), a conical helix
phase (cone), a field-polarized state (FP, $H>H_{c2}$), and a complex A-region
with several pockets. Various crossovers occur in the ($H,T$)-range $T_C < T <
T_0=280$~K and $H\lesssim 0.45$~kOe. The lines $H_{\times}^{<}(T)$ and
$H_{\times}^{>}(T)$ indicate a crossover from the paramagnetic phase (PM) to
the FP state. \label{fig:combinedPhaseDia}}
\end{figure}

The information obtained
from the \chiach and \chiact data is summarized in a
common phase diagram presented in Fig.~\ref{fig:combinedPhaseDia}.
This diagram establishes the existence of a complex A-region in FeGe.
Long-range magnetic order sets in at $T_C=278.2(3)$~K (for $H=0$).
The helical modulation along the $[100]$ direction persists below \hcone
which is of the order of 60~Oe.
For larger external fields the propagation direction of the helix is forced
into the direction along the field and a conical helix is formed. Finally, the
fully field-polarized state is entered at the spin-flip field $H_{c2}$ where
the cone angle closes.
This is the normal behavior of a
cubic helimagnet which is observed in the major temperature
range below $T_C$.
However, the A-region observed between 273~K and 278~K
is found to be split into several distinct pockets.
The main part, designated $A_1$ in Fig.~\ref{fig:combinedPhaseDia}, is clearly
separated by a phase transition from phase $A_2$ which exists at higher fields.
%
%
The magnetic structure of the $A_2$ region, however, appears to transform
continuously into the conical phase, as \chiact for increasing field does not
show clear anomalies above 0.5~kOe.
There are two other small pockets $A_0$ and $A_3$ which
are distinct from $A_1$ and the conical phase.

At high magnetic fields, we observe that the field-aligned state is separated
from the paramagnetic phase by a broad crossover region, indicated by the
dashed lines $H_{\times}^{<}$ and $H_{\times}^{>}$ in
Fig.~\ref{fig:combinedPhaseDia}.
This crossover region transforms below about 0.45~kOe into a region with
precursor states between $T_C$ and $T_0$.
In this region, we have observed smeared anomalies both in $H$- and $T$-runs,
as indicated in the phase diagram by the dashed lines. Moreover, closer to the
A-region there is evidence of further transformations or crossovers that
could not be resolved from the experiments, e.g. related to the shallow anomaly
in \chiach at 278.5~K (Fig.~\ref{fig:chilowtzoom}(b)).
The two important observations from this phase diagram are:
(i) The phases with
magnetic long-range order are separated from the paramagnetic state by a
wide temperature range which we call the precursor state.
(ii) The A-region is composed of several distinct phase pockets.

In theory, the occurrence of an anomalous precursor
regime in chiral magnets near magnetic ordering relies on the formation of
localized (solitonic) states \cite{Roessler06,Leonov10}.
In non-centrosymmetric magnets with DM interactions, the twisting of the
spin-structure energetically favors the formation of
1D helicoidal kinks \cite{Dzyaloshinskii64},
2D Skyrmions \cite{Bogdanov89,Bogdanov94},
or even 3D spherulites.
Theoretical analysis of magnetic states
in a cubic helimagnet can be based
on the standard Dzyaloshinskii model \cite{Dzyaloshinskii64,Bak80}.
A crucial feature of this model has been revealed by analysis of the
interactions between helicoidal kinks \cite{Schaub85,Yamashita85} and Skyrmions
\cite{Leonov10,Roessler10}.
These interactions, being \textit{repulsive} in a broad temperature range,
become attractive in the vicinity of the ordering temperature, the region of
the precursor phenomena.
This range is determined by a characteristic
{\em confinement temperature} $T_L$ \cite{Leonov10}.
Therefore, magnetic ordering
occurs as the simultaneous nucleation and
condensation of stable solitonic units
with attractive interactions.
These phenomena have a universal character
as they derive from the coupling
between angular degrees of freedom of magnetization to its
longitudinal magnitude,
which becomes relevant near the paramagnetic state \cite{Leonov10}.
This mechanism destroys homogeneity of the order parameter.

\begin{figure}
\includegraphics[width=0.425\textwidth]{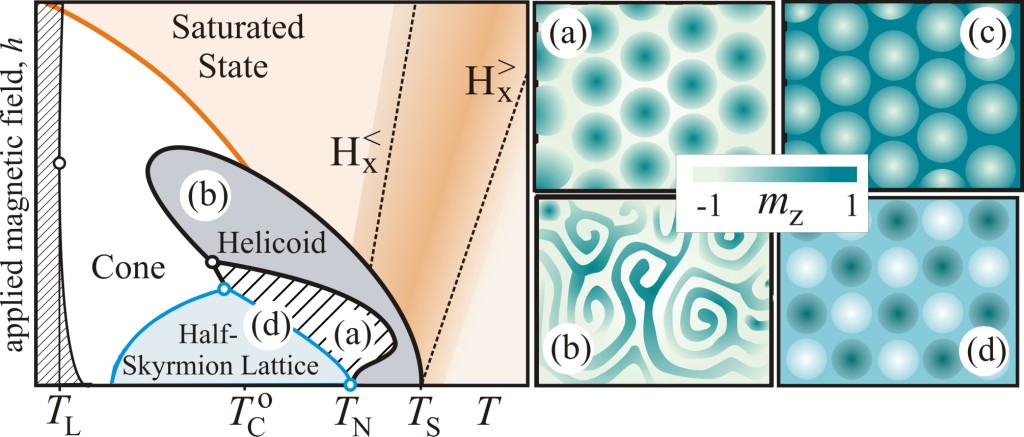}
\caption{%
Sketch of a theoretical phase diagram for chiral magnets
near magnetic ordering according to Ref.~\cite{Leonov10}.
(See text for details.)
In an applied field a densely packed full $+\pi$ Skyrmion lattice is found in
region (a).
The helicoid transverse to an
applied field is reentrant in region (b).
Region (d) is a half-Skyrmion lattice with defects.
Insets are maps of magnetization component $m_z$ for different modulated
magnetic structures in cutting planes perpendicular to the applied field along
$\mathbf{z}$: (a) the dense packed $+\pi$ Skyrmion lattice, (b) spiral-domain
texture of helicoids, adapted from pictures of cholesterics with polygonalizd
helix domains \cite{Bouligand72}, (c) $-\pi$-Skyrmion lattice (occurring well
below $T_L$), and (d) half-Skyrmion square lattice.\label{Fig4Theory} }
\end{figure}

The details of the mesophase formation in a real magnetic system sensitively
depend on very small additional effects such as, e.g., anisotropy, dipolar
interactions or fluctuations.
However, some characteristic features of the experimental phase diagram
depicted in Fig.~\ref{fig:combinedPhaseDia} can be interpreted in the framework
of the modified Dzyaloshinskii model for metallic chiral cubic
helimagnets\cite{Roessler06} that neglects such secondary effects but is able
to provide a more realistic description of the inhomogeneous twisted magnetic
structure in these mesophases.
Figure \ref{Fig4Theory} displays a sketch of the magnetic phase diagram,
derived from the theoretical model of Refs.~\cite{Roessler06,Leonov10}.
Above the confinement temperature $T_L$ chiral
modulations are clustering and form mesophases in the precursor region
$T_L<T<T_S$.
$T_S$ extends well above the ideal Curie temperature $T_C^0$ for
ferromagnetic long-range order in the absence of DM couplings
and $T_N$ for helix magnetic ordering.
Localized chiral modulations become stable at $T_S$ {\em before discernible}
magnetic long-range order is established. 
This is the region of distinct anomalies
between the paramagnetic state and the
helical states, and between the field-polarized state
at higher fields, $H_x^{>}$ and $H_x^{<}$
(see Fig.~\ref{fig:combinedPhaseDia}).
As the formation of a chiral condensate, composed from these localized
entities, is not accompanied by a diverging correlation length, only a smooth
crossover is expected.

The different A-pockets can be identified as partially ordered or crystalline
mesophases that may form below this cross-over region.
The theoretical phase diagram shows that both helicoidal kink-like and
Skyrmionic precursors may exist, which can explain a splitting of the A-phase
into distinct pockets.
Thus, the phase diagram
contains several solitonic equilibrium phases:
a hexagonal $+\pi$-Skyrmion lattice (marked (a) and inset (a)) indicates the
existence of densely packed phases of baby-Skyrmion strings
\cite{Leonov10,Roessler10}. This Skyrmion phase competes with a helicoidal
phase (b) stabilized at higher fields.
In contrast, the $-\pi$-Skyrmion lattice states (c) expected to form a
metastable low-temperature phase in chiral cubic
helimagnets\cite{Bogdanov89,Bogdanov94}, do not exist near magnetic ordering in
this model.
At low fields, a half-Skyrmion lattice (d)
suggests a further competing type of precursor
with split Skyrmionic units and defect points
or lines, where magnetization passes through zero.
At lower temperatures,
long-range ordered condensed phases may finally form.
The observed Curie temperature $T_C$
is expected below $T_N$ owing to the competition
with spontaneous half-Skyrmion states\cite{Roessler06}.
%

A comparison of the experimental phase diagram of FeGe with the theoretical
results suggests to identify the pocket A$_1$ in
Fig.~\ref{fig:combinedPhaseDia} with a densely packed phase consisting of full
$+\pi$ Skyrmions, the region and inset '(a)' in Fig.~\ref{Fig4Theory}.
The pocket A$_2$ at higher fields then should be a reentrant helicoidal state
labeled as '(b)' in Fig.~\ref{Fig4Theory}.
A helicoid can be continuously transformed by
re-orientation of the propagation direction into the conical state.
This conforms with the open region A$_2$ in Fig.~\ref{fig:combinedPhaseDia}.
As we know the A$_2$ region to display characteristic six-fold Bragg patterns
in magnetic neutron diffraction \cite{Wilhelm11}, this area most likely is a
polygonalized helix-domain structure.
Such defect textures
are well known from cholesteric liquid crystals \cite{Bouligand72}
that obey a similar phenomenological theory as chiral magnets.
In these helicoidal textures, domains with nearly circular cross-section
containing spiralling helices are arranged into lattices
or into a hexatic orientational order.
In fact, spiralling domains of the helicoidal spin-structure
have earlier been seen by Lorentz transmission electron microscopy
in FeGe thin films \cite{Uchida08}.

In conclusion, the ac-magnetic susceptibility data on cubic FeGe show a complex
sequence of phase transitions and crossovers in the vicinity of
$T_C=278.2(3)$~K for fields applied parallel to the [100] direction. The
A-phase region is not homogeneous, but split into several parts.
The results exemplify the complex nature of precursor phenomena in cubic
helimagnets with chiral Dzyaloshinskii-Moriya couplings.
The magnetic ordering transition in these materials
is dominated by the formation of localized states
and their organization into mesophases interleaved between paramagnetic and
helical ground states.
The identification of different A-phases
also shows that
understanding of the spin-structures
in the precursor region is incomplete and
that the identification of Skyrmions and
Skyrmion lattices close
to magnetic ordering is an open problem.


We acknowledge fruitful discussions with Yu. Grin and technical assistance from
R. Koban, Yu. Prots, and H. Rave. Support by DFG project RO 2238/9-1 is
gratefully acknowledged.
%

\end{document}